# An FFT based measure of phase synchronization

Kaushik Majumdar

*Odyssee Group*, INRIA Sophia Antipolis, 2004 Route des Lucioles, 06902 Sophia Antipolis, France; E-mail: mkkaushik@hotmail.com

**Abstract**

In this paper phase of a signal has been viewed from a different angle. According to this view a signal can have countably infinitely many phases, one associated with each Fourier component. In other words each frequency has a phase associated with it. It has been shown that if two signals are phase synchronous then the difference between phases at a given component changes very slowly across the subsequent components. This leads to an FFT based phase synchronization measuring algorithm between any two signals. The algorithm does not take any more time than the FFT itself. Mathematical motivations as well as some results of implementation of the algorithm on artificially generated signals and real EEG signals have been presented.

## 1. Introduction

Synchronization is an important concept in neuroscience. However despite its importance in many areas of science including physics and biology there is no universally accepted notion for synchronization either in physics or biology (Mormann et al., 2000). In this paper we will be concerned about phase synchronization between any two signals irregardless of the power. There is no unique method to extract phase of a general signal (Kreuz et al., 2007). A comparison of many existing methods can be found in (Bruns, 2004) and (Le Van Quyen et al., 2001). In this paper we are going to propose a new, rather straight forward and fast method to measure phase synchronization between any two signals. The method is based on FFT and does not take any more time than the FFT itself. In section 2 we will give some motivations. In section 3 we will introduce the new measure of phase synchronization. In section 4 we will present an algorithm to calculate the phase synchronization. In section 5 we will present some results on simulated signals as well as human scalp EEG signals. We will conclude the paper with a section on discussion and future directions.

## 2. Phase synchronization

In the Figure 2.1 both the pair of signals must be phase synchronous, because both the pairs are actually superposition of the same signal with itself only after contracting the amplitude by two different factors. Geometrically the phase synchronization in the top pair looks rather obvious, but in the bottom pair it is not that obvious. But if the top pair and the bottom pair should both be phase synchronous there must be some factor which does not change from one pair to the other. In this paper we are going to investigate what this factor may be.

Usually a signal is defined by its amplitude as a function of time. In Figure 2.1 we see how the appearance of a signal can vary dramatically if we uniformly contract or blow up the amplitude. By contracting or expanding the amplitude the phase is not at all affected and therefore when we are interested in the phase alone we can define a signal by its phase as a function of time instead. Let two signals be represented by $\phi_1(t)$ and $\phi_2(t)$, where $\phi_i(t), i=1,2$ are the phases of the signals and $t$ is the time. The dynamics of the evolution of the two oscillators or signals with single frequency can be represented by the following two equations when there is some coupling between the two (Rosenblum et al., 2001):

$$\frac{d\phi_1}{dt} = \omega_1 + \epsilon\, g_1(\phi_1, \phi_2) \tag{2.1}$$

and

$$\frac{d\phi_2}{dt} = \omega_2 + \epsilon\, g_2(\phi_2, \phi_1), \tag{2.2}$$

where $\omega_1$, $\omega_2$ are single frequencies associated one with each signal or oscillator, $\epsilon$ is the coupling constant and $g_1$, $g_2$ are coupling functions $2\pi$ periodic in both $\phi_1$ and $\phi_2$.

The interaction between the signals essentially affect the evolution of their phases if the frequencies $\omega_1$ and $\omega_2$ are in resonance, which implies

$$n\omega_1 \approx m\omega_2, \tag{2.3}$$

where $n, m$ are integers. (2.1), (2.2) and (2.3) together imply

$$\frac{d(n\phi_1 - m\phi_2)}{dt} \approx \epsilon\, (ng_1 - mg_2). \tag{2.4}$$

By adjusting $\epsilon$ we can make (2.4) as

$$\frac{d(n\phi_1 - m\phi_2)}{dt} = \epsilon\, (ng_1 - mg_2). \tag{2.5}$$

(2.5) is an ordinary linear differential equation whose phase space can only have a fixed point or a limit cycle. In case of an attractive fixed point the left hand side of (2.5) becomes zero leading to

$$n\phi_1 - m\phi_2 = C, \tag{2.6}$$

where $C$ is a constant. (2.6) is the condition for perfect phase locking (Rosenblum et al., 2001), which leads to

$$ng_1 = mg_2. \tag{2.7}$$

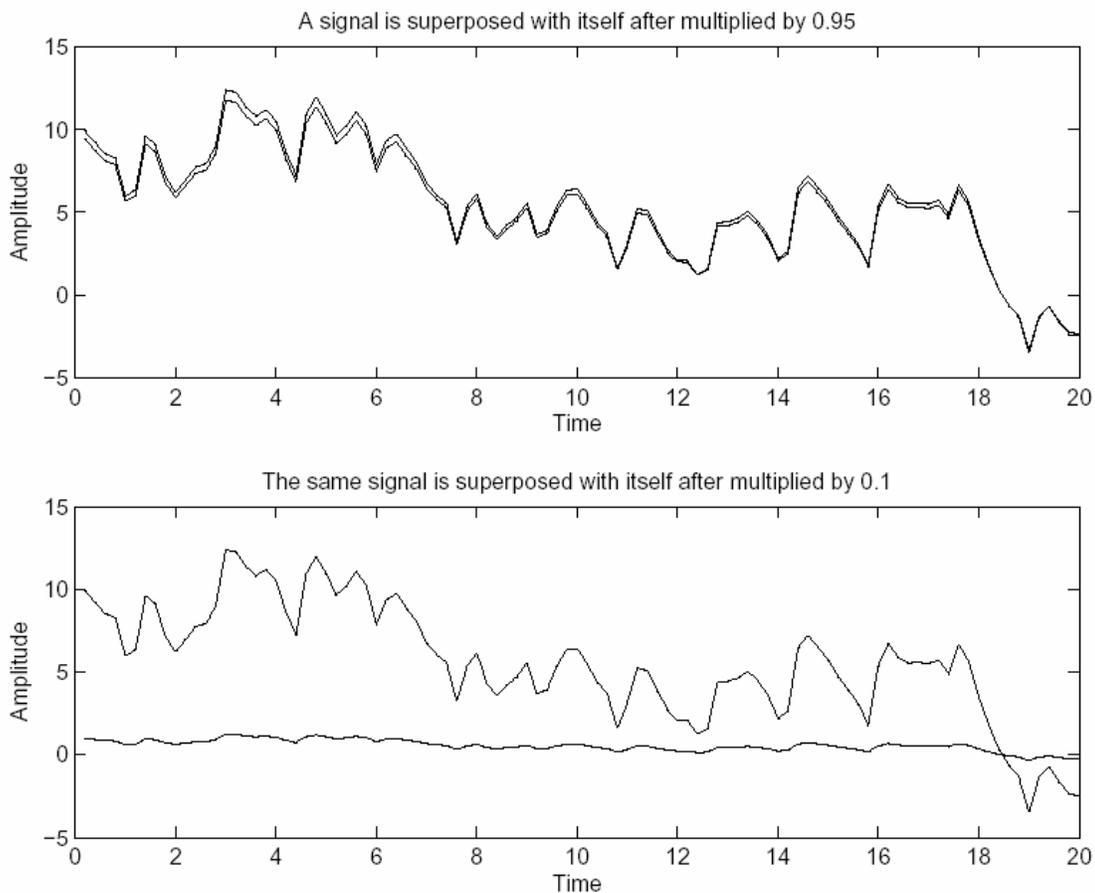

Figure 2.1. Both the human scalp EEG signal pairs in the top as well as in the bottom must be phase synchronous, because only their amplitude has been contracted by different factors.

Let us consider the simplest case when $\omega_1 = \omega_2$. Then (2.3) implies $n = m$ and by (2.7) we have

$$g_1 = g_2. \tag{2.8}$$

In the first order approximation it becomes

$$g_1(\phi_1,\phi_2) = g_2(\phi_2,\phi_1) \approx |\phi_1 - \phi_2|. \tag{2.9}$$

## 3. Measuring the synchronization

Let $x_j(t)$ and $x_k(t)$ be any two signals. Their Fourier expansions can be written as following:

$$x_j(t) = \frac{a_{j0}}{2} + \sum_{n=1}^{\infty}(a_{jn}^2 + b_{jn}^2)^{1/2}\sin\left(\frac{2\pi nt}{p} + \alpha_{jn}\right) \tag{3.1}$$

$$\alpha_{jn} = \tan^{-1}\left(\frac{a_{jn}}{b_{jn}}\right) \tag{3.2}$$

$$x_k(t) = \frac{a_{k0}}{2} + \sum_{n=1}^{\infty}(a_{kn}^2 + b_{kn}^2)^{1/2}\sin\left(\frac{2\pi nt}{p} + \alpha_{kn}\right) \tag{3.3}$$

$$\alpha_{kn} = \tan^{-1}\left(\frac{a_{kn}}{b_{kn}}\right) \tag{3.4}$$

It is clear from (3.1) and (3.3) that the general signals can be treated as superposition of simplest or fundamental signals generated by oscillators with constant frequency and phase. If the two signals are to be perfectly phase synchronous then the generating oscillators with $n$ th frequency for both the signals must have same phase, which literally means

$$\alpha_{jn} = \alpha_{kn}. \tag{3.5}$$

However we are interested in a weaker condition in which two signals are not perfectly, but approximately phase synchronous. In that case we can relax (3.5) as

$$|\alpha_{jn} - \alpha_{kn}| \approx 0. \tag{3.6}$$

Notice that we are interested purely in phase synchronization between two signals irregardless of the amplitude of the fundamental oscillators. The power spectral density estimate of the two perfectly phase synchronous EEG signals in the second subplot of Figure 2.1 has been presented in Figure 3.1. Even when the power spectral density comes close to zero it contains valuable information about the phase synchronization. Therefore in order to get an accurate measure of phase synchronization (or phase asynchronization) it is essential to consider $|\alpha_{jn} - \alpha_{kn}|$ for all $n$, irregardless of the power associated. For a pair of perfectly phase synchronous signals $|\alpha_{jn} - \alpha_{kn}|$ should be zero for all $n$ and for closely phase synchronous signals $|\alpha_{jn} - \alpha_{kn}| \approx 0$ should hold for all $n$. This implies that

$$syn(x_j(t), x_k(t)) = mean(|\alpha_{jn} - \alpha_{kn}|) + std(|\alpha_{jn} - \alpha_{kn}|) \tag{3.7}$$

should be a small quantity, while mean and standard deviation ($std$) has been taken across all $n$.

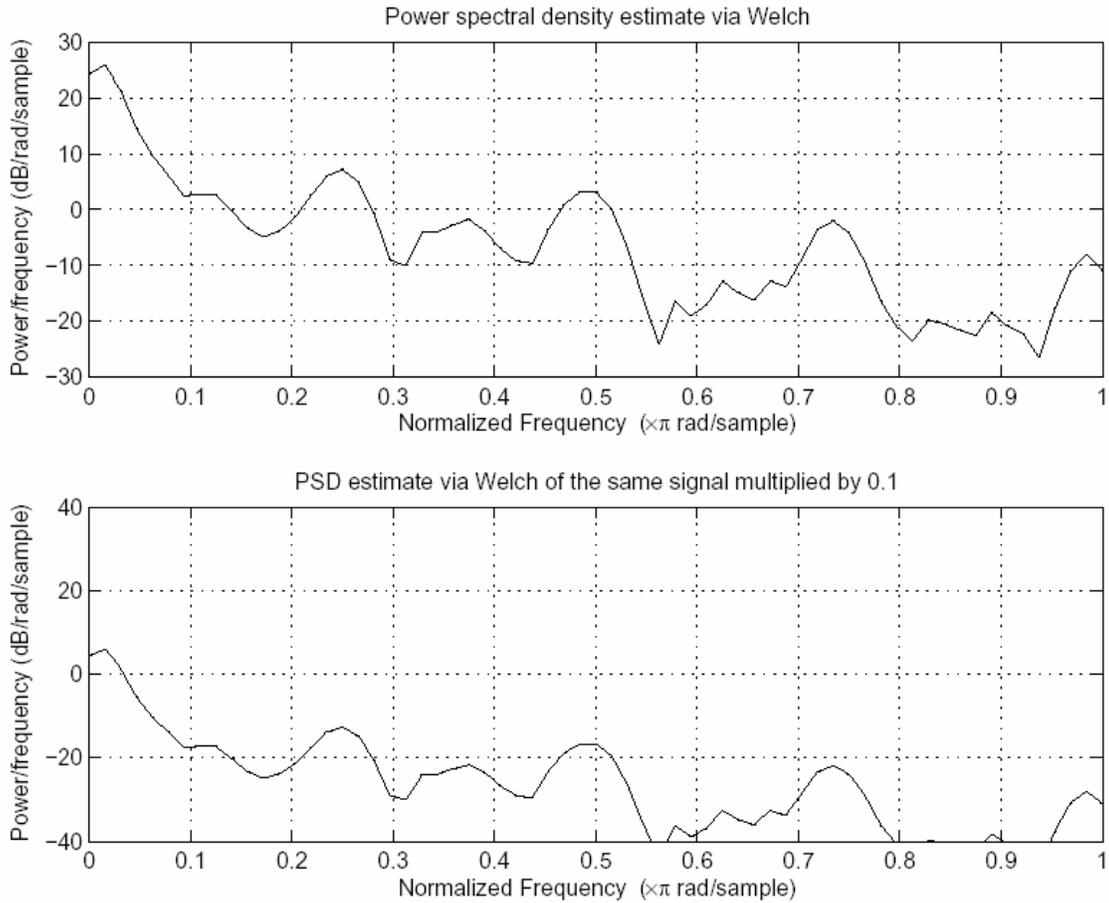

Figure 3.1. Power spectral density estimate of the two EEG signals that have been shown in the second subplot of Figure 2.1.

The *syn* function defined on a pair of signals as given by (3.7) is a quantity which remains invariant across pairs of signals with same degree of phase synchronization, for example it is equal to zero for a pair of perfectly phase synchronous signals. This has been tested with artificially generated EEG signals on a real head model of a human subject. The head model consists of cortex, skull and scalp. All three surfaces have been extracted from the subject's structural MRI data. The cortical surface consists of a triangular mesh constructed with 8959 points. There are 60 electrode positions on the scalp according to the international 10/10 system (also recorded by the same MRI), through which EEG has to be collected. The forward calculation on this head model has been performed by the boundary element method (Kybic et al., 2005), with the help of an open source software OPENMEEG (OPENMEEG, 2007) developed in our lab. A single source has been constructed by activating 5 closely contiguous cortical points with an artificial signal consisting of a few arbitrary Fourier components. For the generated EEG at the 60 scalp channels the value of the *syn* function for each pair as given by (3.7) has exactly been 0. The simulation has been repeated with different activating signals for the source and different source positions with the same result every time.

**4. The algorithm**

(3.6) together with (3.2) and (3.4) implies

$$\left| \frac{a_{jn}b_{kn} - a_{kn}b_{jn}}{a_{jn}a_{kn} + b_{jn}b_{kn}} \right| \approx 0, \qquad (4.1)$$

for all $n$. In particular

$$E(n) = \left| \frac{a_{jn}b_{kn} - a_{kn}b_{jn}}{a_{jn}a_{kn} + b_{jn}b_{kn}} - \frac{a_{jn+1}b_{kn+1} - a_{kn+1}b_{jn+1}}{a_{jn+1}a_{kn+1} + b_{jn+1}b_{kn+1}} \right| \qquad (4.2)$$

are very small quantities when $x_i(t), x_j(t)$ are almost phase synchronous, and hence the mean and the standard deviation of $E(n)$ across all $n$ will also be a very small quantity. The following pseudo code will return the normalized measure of phase synchronization between two signals.

**Proc(synchronization_detection)**
**Input:** $x_j(t), x_k(t), p$ ;
**Output:** $syn(x_j(t), x_k(t))$ ;
  1. **A** ← **FFT**($x_j(t), [0, p]$);
  2. **B** ← **FFT**($x_k(t), [0, p]$);
  3. **for** ($1 \leq i \leq \left\lceil \frac{p}{2} \right\rceil$)
    $Z[i] \leftarrow real(A[i])real(B[i]) + imag(A[i])imag(B[i])$ ;
        **if**($Z[i] \neq 0$)
            $D[i] \leftarrow \dfrac{real(A[i])imag(B[i]) - real(B[i])imag(A[i])}{Z[i]}$ ;
    **else**

        **print("Signals are asynchronous.");**

4. **for** $(1 \leq i \leq \lceil \frac{p}{2} \rceil - 1)$

        $E[i] \leftarrow |D[i+1] - D[i]|$;

5. $syn(x_j(t), x_k(t)) = (1 + mean(E(i)) + std(E(i)))^{-1}$;

**return** $syn(x_j(t), x_k(t))$;

*imag* stands for imaginary. It is essential that $Z[i]$ does not become zero. The thousands of pairs of signals we have worked with we never encountered $Z[i] = 0$ for any $i$ (Majumdar, 2008). In case $Z[i] = 0$ holds for some $i$, then it is clear from (2) and (4) that the *ith* harmonics of $x_j(t)$ and $x_k(t)$ have phase difference equal to $\pi/2$ and the signals cannot be synchronous. It is easy to check that the time complexity of the above algorithm equals to that of the FFT (Majumdar, 2006).

Notice that to make the measure of phase synchronization normalized we have taken

$$syn(x_j(t), x_k(t)) = (1 + mean(E(i)) + std(E(i)))^{-1}. \tag{4.3}$$

The more phase synchronous the signals are the less is the quantity $mean(E(i)) + std(E(i))$. In order to make the *syn* as an increasing function rather than decreasing with respect to phase synchronization we have taken the reciprocal. $mean(E(i)) + std(E(i))$ is zero when $x_i(t) = x_j(t)$. But then the signals are perfectly phase synchronous. In order to assign the highest value to the *syn* function in those cases in the 0 to 1 scale we have chosen to add 1 before doing the inversion. Apart from this, adding 1 also helps to resolve some numerical artifacts introduced by the computer due to truncating floating point numbers.

## 5. Results

In this section we will be presenting the results of some simulation studies to show the effectiveness of the algorithm described in the last section. First we have constructed ten artificial signals by adding a few arbitrary Fourier components for each of them. Only two of them are phase synchronous with each other. No other pair of signals are phase synchronous. When the above algorithm was run on each pair (45 in total) only for the synchronous pair the result was 1. For all the other 44 it was less than or equal to 0.0881.

Extensive simulations have been done on a real head model constructed out of the subject's structural MRI data. The aim of the simulations were to estimate location of the cortical sources (with known position) from the scalp EEG data (generated by the forward calculation) with the help of the phase synchronization and signal power profile in the neighborhood of each channel. This has been compared with source localization by classical minimum $L_2$ norm inverse method with excellent compatibility. Then the method was applied to estimate the cortical sources from scalp EEG data (within a 4.5 cm of error margin as measured on the scalp) of the subject during median nerve stimulation. The detailed results have been presented in (Majumdar, 2008). Here we will present three diagrams of human scalp EEG pairs and the corresponding phase synchronization values calculated by the above algorithm.

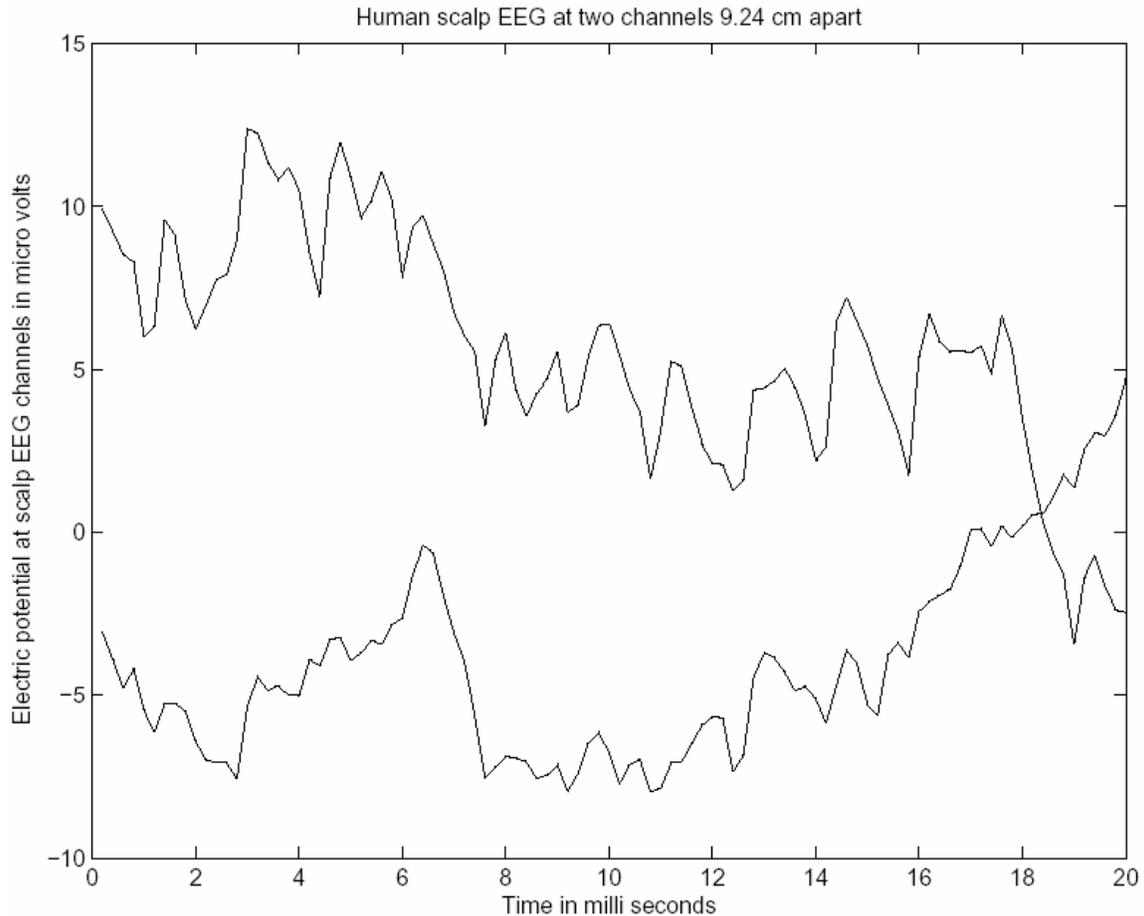

Figure 5.1. The value of phase synchronization between the two EEG signals is 0.0001 in 0 to 1 scale. Closeness of the value to 0 signifies high degree of asynchrony between the two signals, which is also evident from their time vs. amplitude plot.

The proposed measure of phase synchronization is very sensitive to even small changes, which we have observed at the time of running the algorithm on human EEG data. The phase synchronization value between any two channels often varies significantly from one trial to the next. Marked variations have been observed in most cases even if the time window is slid just by 1 millisecond, whereas the signal power profile remains almost identical after the sliding. We have worked in most cases with 12 ms long windows on a 5000 Hz sample frequency data set. The algorithm is indeed sensitive to noise, but we have chosen sensitivity over stability in order to obtain more minute information whenever possible. The popular belief that closer the two channels are the higher is the phase synchronization value because of the volume conduction is not true in general. In many cases far away channels behaved more synchronously than closer channel pairs during particular trials. Signal pairs in Figure 5.1 and Figure 5.2 are both highly asynchronous, but their degree of asynchrony is not the same, which is not so obvious from the time vs. amplitude plot.

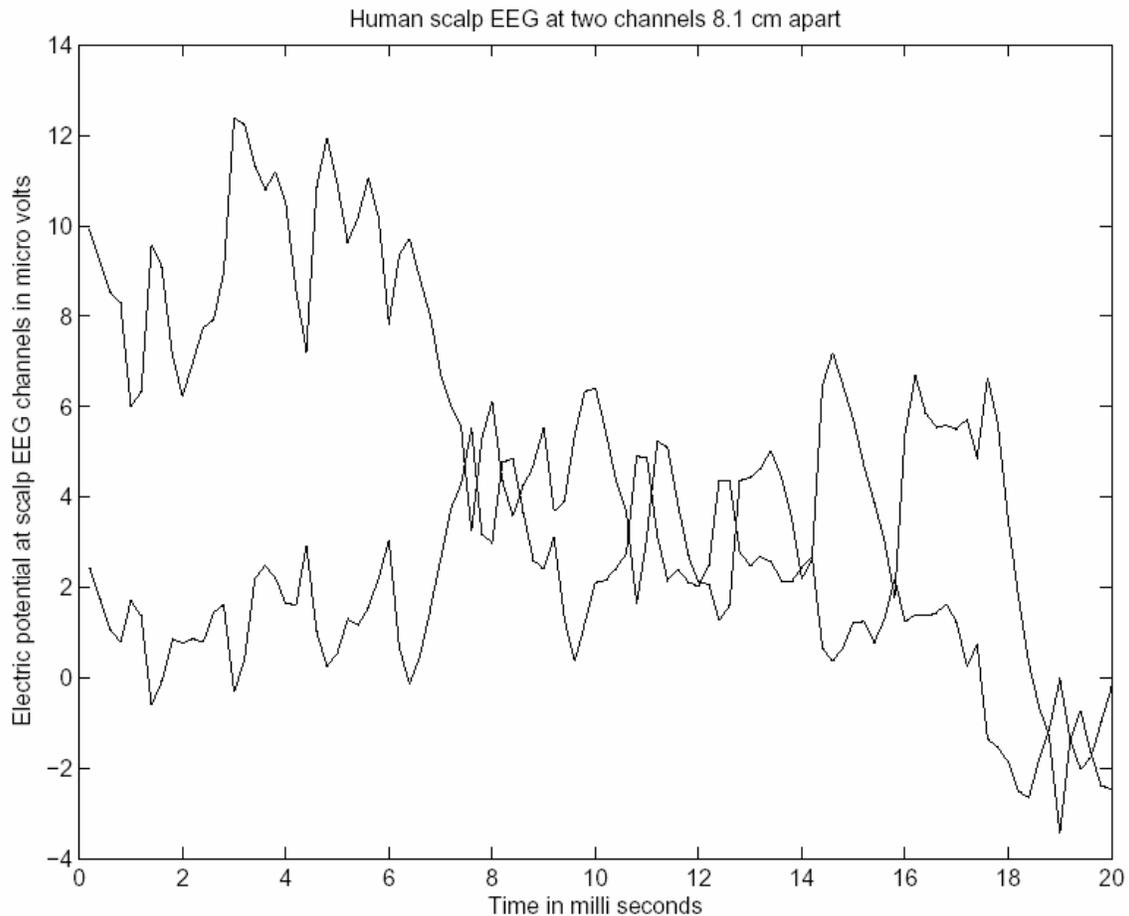

Figure 5.2. The value of phase synchronization between the two EEG signals is 0.0048 in 0 to 1 scale. The signal pair is highly asynchronous as evident from their time vs. amplitude plot.

Contrary to the signal pairs in Figure 5.1 and Figure 5.2 the signal pair in Figure 5.3 are highly phase synchronous. Probably the signal pair in Figure 5.3 looks more phase synchronous than the signal pair in the second subplot of Figure 2.1, which is not true. Because the value of phase synchronization in the latter case is 1 and in the former it is 0.7759.

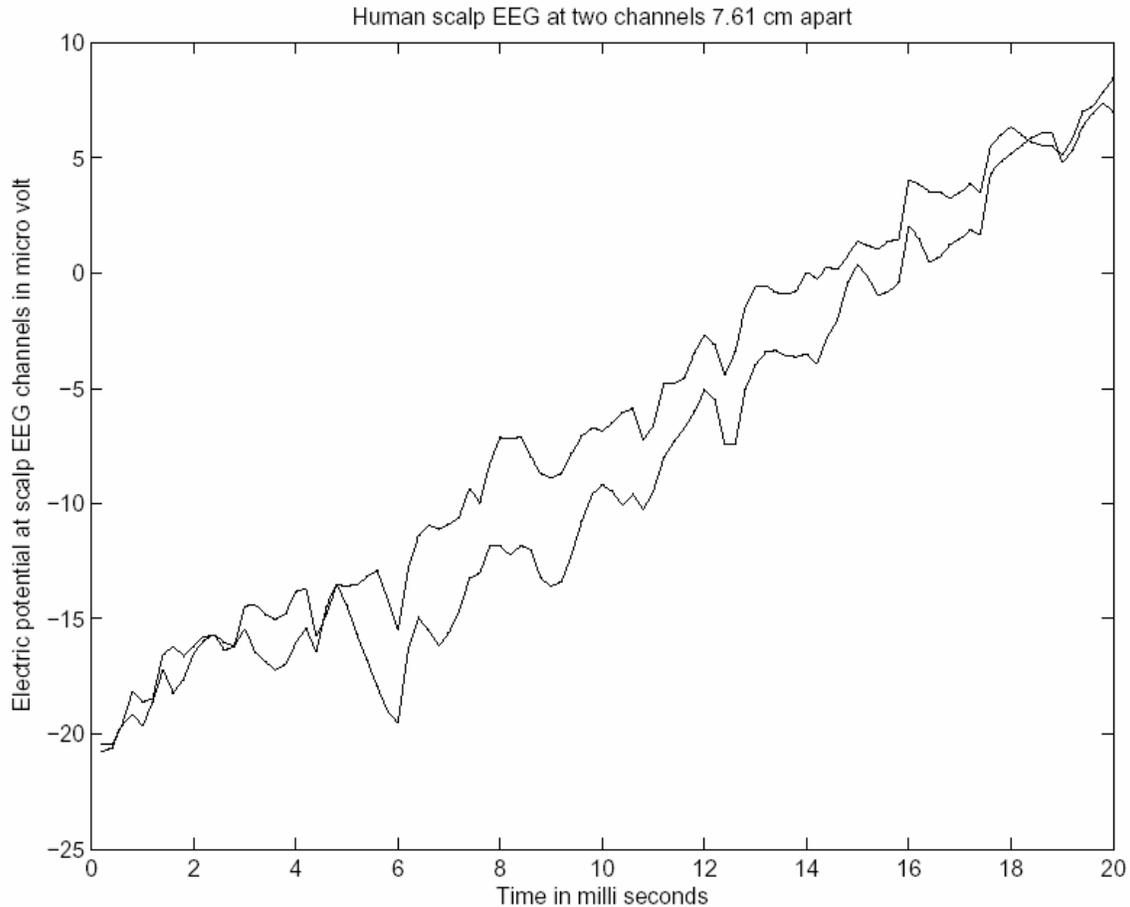

Figure 5.3. The value of phase synchronization between the two EEG signals is 0.7759 in 0 to 1 scale. The signal pair is highly phase synchronous as evident from their time vs. amplitude plot.

**Discussion and future directions**

In this paper we have proposed a new measure of phase synchronization between a pair of signals. We have presented mathematical arguments as well as simulation results on real and artificial data to substantiate the new measure. However we have not presented any comparative study vis-à-vis any of the existing phase synchronization measuring algorithm. This is because there is no single method for detection of phase synchronization. Instead different methods are applied for different systems. We have done extensive simulations and have found that our method is working pretty well for human scalp electrophysiological data as well as on synthetic data. Many of those results have been presented in (Majumdar, 2008) from a cortical source localization point of view. In this paper we have tried to present phase synchronization in a balanced way divided between theory and practice.

Since the algorithm is very fast (time complexity is $n \log n$, where $n$ is the input size or the number of time points in the pair of signals) it can efficiently be used on sliding time window with good temporal resolution, which in many cases may make it as efficient as wavelet based algorithms (Bruns, 2004). As our discussions made it clear that the sensitivity of the algorithm can produce quite interestingly discernable results on functional scalp electrophysiological data and therefore it can be used to non-invasively study effects of perception like the one carried out in (Rodriguez et al., 1999). The results in (Majumdar, 2008) clearly indicates that this can be done even for single trials, unlike in (Rodriguez et al., 1999), where only the average across trials has been considered. Importance of the study of phase synchronization in the single trial electrophysiological signals has been well recognized in (Baillet et al., 2001). It can probably find quite useful applications in epilepsy research, particularly in seizure focus lateralization (Caparos et al., 2005).

One can also take $\left|\alpha_{jn} - \alpha_{kn}\right|$ only for selective $n$, according to associated power etc. In that case the algorithm with slight modification will be able to determine power synchronization or some other form of coherence between a pair of signals.

**Acknowledgement**

The current work has partly been supported by an EADS grant under the contract no. 2118. Maureen Clerc, Theodore Papadopoulo and Olivier Faugeras are being acknowledged for discussions and many helpful suggestions. Patrick Marquis is being acknowledged for providing the human EEG data.